# Complex dynamics and development of behavioural individuality

David N. Fisher*, Matthew Brachmann, Joseph B. Burant

Department of Integrative Biology, University of Guelph, Guelph, Ontario, Canada

*contact: davidnfisher@hotmail.com

## Main text

### *Consistently finding consistent differences*

Animals in a population frequently display consistent among-individual differences in their behaviours (Bell, Hankison, & Laskowski, 2009; D. S. Wilson, 1998). Among-individual differences in behaviour have long been considered important as such differences are required for natural selection to cause mean individual behaviour to change. Indeed, in the last few decades increased interest in these differences has spurred the generation of the field of "animal personality". A key element of this recent surge has been to identify the processes that lead to and maintain among-individual behavioural differences. Correlations with more stable physiological or motivational states (Sih et al., 2014) or life-history traits (Réale et al., 2010), or the role of environmental factors (Kortet, Hedrick, & Vainikka, 2010; Montiglio & Royauté, 2014) have all been investigated to explain why animals in the same population show consistent differences in mean behaviour across contexts (Dingemanse & Wolf, 2010; Kight, David, & Dall, 2013). These explanations typically require there to be variation in some other factor, be that genetic or environmental in origin, which then drives among-individual differences in behaviour (but see: Luttbeg & Sih, 2010; Mathot, Dekinga, & Piersma, 2017; Sih et al., 2014).

Recent research, however, has found evidence for differences among-individuals, of the magnitude very commonly found in nature (Bell et al., 2009), in situations where any obvious variation of other factors was lacking. Most recently, Bierbach et al. (2017) found that among-individual differences in behaviour occur in Amazon mollies (*Poecilia formosa*) that did not differ in terms of genetic or environmental background. This was achieved by using clonal fish reared in identical tanks to reduce genetic and environmental variation. Similar results have been achieved in inbred eastern mosquitofish (*Gambusia holbrooki*) (Polverino, Cigliano, Nakayama, & Mehner, 2016) and inbred mice (*Mus musculus*) (Brust, Schindler, & Lewejohann, 2015; Freund et al., 2013), although some degree of genetic variation is likely to persist in these latter examples. The authors suggested that finding consistent among-individual differences in behaviour in the absence of any measurable genetic and environmental variation is an unexpected result. This perception is based on an additive linear model of phenotypic variation, where phenotypic variation is the additive sum of genetic variation, environmental variation, and gene-by-environment interactions. Under this paradigm, there is little room for consistent variation among-individuals which is not related to variation in genotypes or the environment. Yet, this is exactly what the authors above found. So how did these animals come to show consistent behavioural differences?

Bierbach et al. (2017) raise a number of potential explanations for the occurrence of individuality against common genetic and environmental backgrounds, including the intriguing possibility of maternal "bet-hedging" through epigenetic variation (see also: Groothuis & Trillmich 2011), and positive feedback between state and behaviours (see also: Sih et al., 2014). However, we would like to advocate an alternative explanation: that among-individual differences may arise regardless of similarity in genetic and environmental background, due to behavioural development being influenced by chaotic dynamics. Bierbach et al. (2017) concluded that individuality may be an inherently unpredictable phenomenon, but hard to predict phenomena that appear stochastic can often be driven by underlying deterministic forces (May, 1976). Weather systems are a good example of this, as they are driven by the deterministic dynamics of air movement, yet are difficult to predict accurately more than a few days or weeks in advance. Dynamics that are sensitive to initial conditions, and so are unpredictable, yet are unpinned by deterministic rules are often called chaotic dynamics.

*Chaotic dynamics*

Reduced long-term predictability for systems with chaotic dynamics hinge on non-linear relationships, where small initial differences in the parameters of the system can lead to much greater differences over time (Boyce, 1992; Hastings, Hom, Ellner, Turchin, & Godfray, 1993). Note that not all non-linear systems are sensitive to initial conditions; some can show convergence to stable "attractors" regardless of initial conditions (Northrop, 2011). An example of a system driven by deterministic chaos, that gives divergent results from small initial differences is a Lorenz attractor. A Lorenz attractor describes patterns of flow (fluid or air) around three "saddle points" in three dimensions (Lorenz, 1963). Individual trajectories circle around one saddle point before flipping over to another, and so on (Fig. 1). Two points that are initially adjacent to one another will rapidly diverge along different trajectories. Therefore, any minute variation between two initial points will lead to potentially substantial differences between them at later time intervals, while any measurement error of initial conditions (which is virtually unavoidable in real-world systems) will render long-term predictions inaccurate. The Lorenz attractor follows the three differential equations below for the movement of the trajectories in three dimensions, based on a continuous time model:

$$\dot{x} = \sigma(y - x)$$
$$\dot{y} = x(\rho - z) - y \quad \quad \quad \text{Equation 1}$$
$$\dot{z} = xy - \beta z$$

where $x$, $y$ and $z$ represent each of three dimensions and $\sigma, \rho$ and $\beta$ are constants in the model. For certain values of these constants the result oscillates, giving chaotic dynamics (Sparrow, 1982). The saddle points act as attractors so that the trajectories are kept within certain boundaries, rather than expanding into infinite space. While this may not necessarily be a suitable model for to describe how consistent differences among-individuals arise, since there are three axes that are continually varying, it is a good model for demonstrating how simple conditions and minimal initial variation can generate large amounts of variation over time.

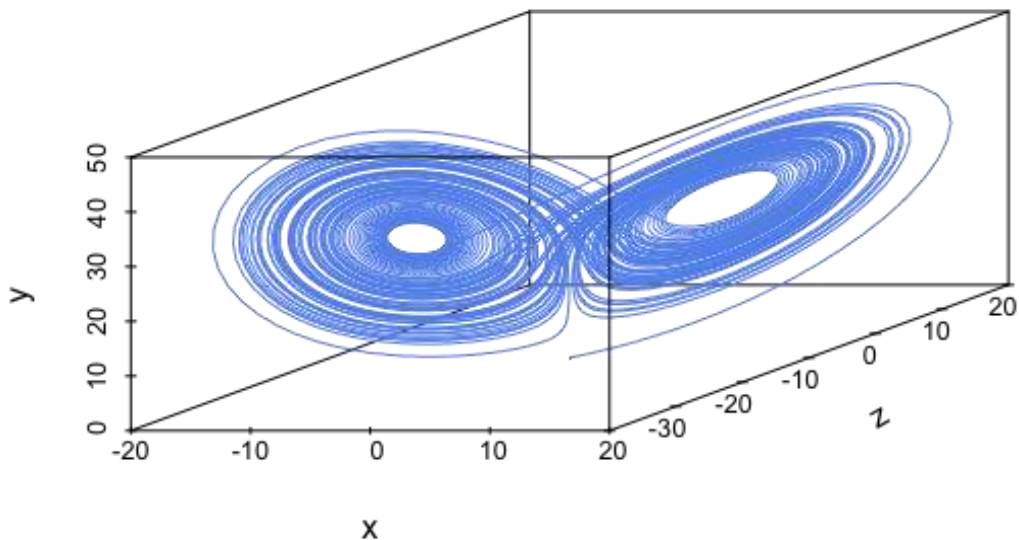

Figure 1. Near-identical initial conditions can lead to divergent locations over time. A Lorenz attractor, developed in the context of air and fluid dynamics, is a relatively simple mathematical model of how non-linear relationships produce seemingly stochastic outcomes in atmospheric convection. Lines represent trajectories of air or fluids. Figure reproduced in R v.3.4.0 (R Development Core Team, 2016) with code modified from <https://gist.github.com/RStyleNinja>. ($\rho = 26.48, \sigma = 10, \beta = 8/3$).

The Lorenz attractor above describes changes in the state of the system in continuous time; similar chaotic dynamics can also be represented in discrete time models. An example of such discrete models is a simple dynamical growth model:

$$x_{t+1} = rx_t(1 - x_t) \qquad \text{Equation 2}$$

where the value of $x$, bounded between 0 and 1, at time $t + 1$ depends on the value at the current time $t$, multiplied by the intrinsic rate of increase $r$, and how far the value is from the maximum (Verhulst, 1838, 1845). Following eq. 2, $r$ values below 1 give extinctions (Fig. 2a), values between 1 and 3.56995 give a single stable value (Fig. 2b) or oscillations among a set of values (Fig. 2c), regardless of the starting population size. Values over 3.56995 however give chaotic dynamics, where no stable value is reached, and small initial differences in the starting values give very different results over time (Fig. 2d).

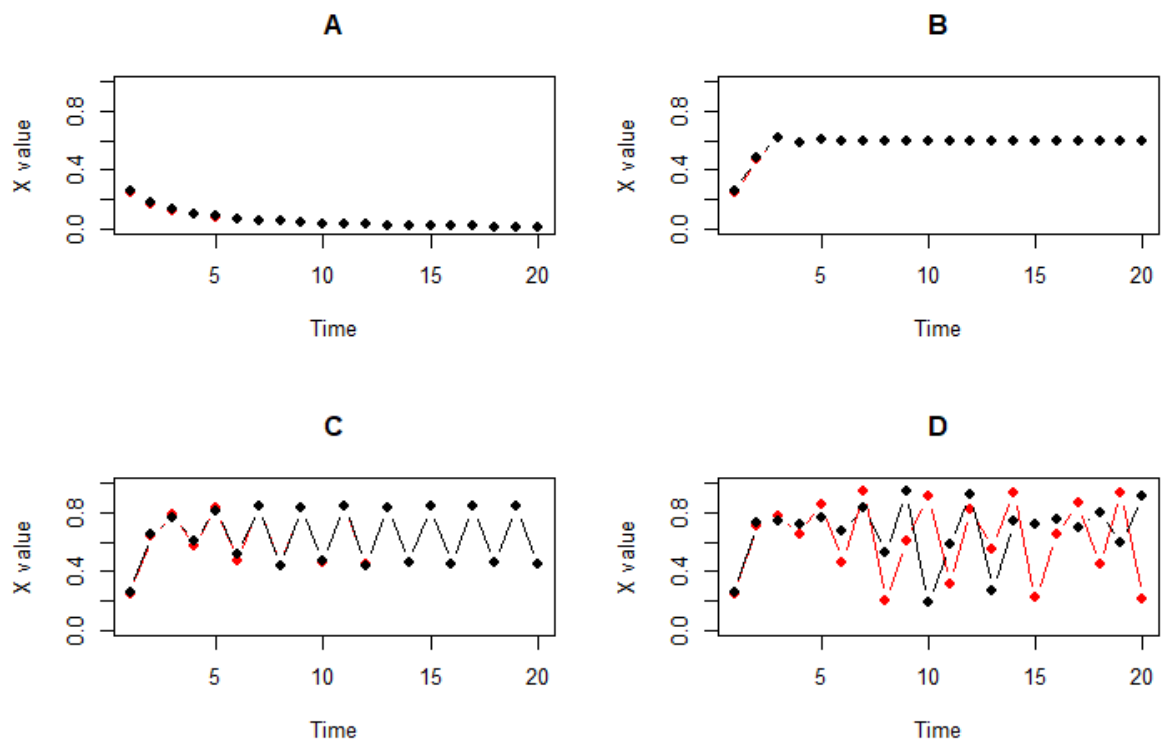

Figure 2. Plots of dynamics of $x$ values over time for different r values following eq. 2. Black lines have a starting value of 0.25, red lines have a starting values of 0.26. Values of $r$ = 0.9, 2.5, 3.4 and 3.8 for A, B, C and D respectively. Once r values go above 3.56995 then no stable state is reached, and trajectories depend on initial values (compare black and red lines in plot D). Figure reproduced in R v.3.4.0 (R Development Core Team, 2016) with code modified from <http://www.magesblog.com/2012/03/logistic-map-feigenbaum-diagram.html>.

With respect to animal behaviour, one could consider $x$ to be the level of the behaviour and $t$ to represent the age of the organism being monitored. Small environmental, genetic, developmental, or stochastic sources of initial differences would then lead to different trajectories for each individual (Fig. 2d). If these different trajectories were then translated into different mean levels of behaviour, and within-individual variation was not especially large, then we would detect significant among-individual variation. The trajectories are prevented from expanding into infinite parameter space by the negative frequency-dependent change inherent in eq. 2. While neither the differential equations (eq. 1) or the difference equations (eq. 2) may be directly mappable to individual behaviour development, they may be useful starting points for understanding how chaos can cause small initial differences to lead to greater differences over time.

If behaviour was chaotic, with trajectories moving unpredictably around in phenotype space, how would consistent among-individual differences be maintained? A simple answer is that behavioural variation may be chaotic during early development, but this chaotic period may end upon maturity, allowing individual trajectories to be maintained. This is perhaps similar to the way that young animals tend to imprint on or copy from conspecifics at an early age, but upon maturity this ceases (Bateson, 1966; Immelmann, 1975). Following maturity, consistent individual behaviours may be maintained at the same level by reinforcement or simply by the costs of plasticity leading to its suppression, giving reduced within-individual variation and so stable differences among-adults (Brust et al., 2015; Duckworth, 2010; Luttbeg & Sih, 2010; Roberts & Del Vecchio, 2000). This has been shown with the house mouse (*Mus musculus*) where individual differences in behaviour are present among inbred individuals reared in identical environments (Brust et al., 2015; Freund et al., 2013). Behaviours tend to be most repeatable in adulthood, with stimuli early in life more likely to influence behaviour than equivalent stimuli later in life (Brust et al., 2015). Similarly, Müller and Müller (2015) found that levels of activity and boldness were not repeatable in mustard leaf beetle (*Phaedon cochleariae*) larvae, but were so among adults. These studies are in line with the suggestion that chaotic, and so variable, dynamics in early development give way to stable behaviours in adulthood.

In some situations, there will be selection for reduced phenotypic variation, such that among–individual variation could theoretically approach zero (Siegal & Bergman, 2002). However, clearly most of the time this selection is not strong enough to overcome whatever drives the formation of personality, as among-individual differences are exceptionally prevalent (Bell et al., 2009). Note that the framework based on an additive linear model leaves no requirement for processes that suppress among-individual variation, as that is the null expectation.

*Non-linearity & Complex systems*

Non-linear models may be useful starting points for understanding the development of behavioural variation. Rather than assuming phenotypes derive from linear, additive relationships, among-individual differences may be the result of non-linear, multiplicative relationships that occur during development. Small initial differences among individuals, from environmental, genetic, developmental, or stochastic sources, could give rise to greater among-individual phenotypic differences in adulthood. This model does not require among-individual differences in states, life-history strategies, or responses to or experiences of environments. Maye et al. (2007) have demonstrated that a degree of genuine stochasticity may well be inherent to animal behaviour, indicating that variation from which among-individual differences may arise is always likely to be present. That the developmental process is key to the emergence of personality differences has been suggested previously (Duckworth, 2010; Groothuis & Trillmich, 2011; Stamps, 2003; Stamps & Groothuis, 2010; Trillmich & Hudson, 2011), but the role of chaotic dynamics specifically has not.

Groothuis and Trillmich (2011) point out both how similar genetic information can lead to different behaviours (e.g., Ayroles et al., 2015), while different genes can produce the same behaviours through canalisation (Siegal & Bergman, 2002). Variation in regulation and expression of the DNA sequence may be more relevant to behaviour than variation in the DNA sequence itself (Groothuis & Trillmich, 2011), hence the DNA sequence similarity of the fish of Bierbach et al. (2017) and Polverino et al. (2016), and the mice of Freund et al. (2013) and Brust et al. (2015), may be considered of minor relevance to the question of whether they will show behavioural differences. This further compound the issue of assuming that behavioural variation is based on standing genetic variation. Those authors do note these points, but we feel it is worth amplifying for the consideration of all researchers working in animal behaviour.

The notion that important differences can arise from equivalent starting points due to chaotic dynamics has received greater consideration outside of animal personality, and outside the behavioural ecology literature generally. For example, significant variation in structure and molecular content has been documented in genetically and environmentally identical cells (Solé & Goodwin, 2000). *Escherichia coli* cells cultured in uniform conditions with uniform genomes developed differences in enzyme activity and grew to colonies of different sizes (Ko, Yomo, & Urabe, 1994). The end-product requires an interplay between the factors involved in cellular activities, not just the information provided by the genes (Solé & Goodwin, 2000).

Such non-additive approaches are more commonly taken in the study of complex systems, where interactions among units of a system, non-linear relationships, and emergent phenomenon are an accepted part of the research paradigm (Bradbury & Vehrencamp, 2014; Hastings et al., 1993; May & Oster, 1976; Solé & Goodwin, 2000). Bradbury and Vehrencamp (2014) called for behavioural ecologists to start incorporating ideas from complexity theory into their research, yet this does not appear to be occurring. Multiplicative relationships often govern animal populations (Boyce, 1992), illustrating the need to consider multiplicative models alongside additive models. For example, when fitting a typical quantitative genetic model (i.e. an animal model; Kruuk, 2004; A. J. Wilson et al., 2010) to behavioural data where individuals showed spontaneous divergence, a large amount of variance would be attributed to permanent environment effects, rather than the additive genetic effect or other effects. This is because individuals would be different from each other, but not for reasons of genetics or any other measured factors. This however does not give much (if any) insight into the mechanisms generating behaviour differences, and furthermore does not explain why behaviour may diverge among-individuals over time. Results such as those presented by Bierbach et al. (2017) provide an ideal opportunity to start thinking a little outside of the box and consider alternative models for behavioural variation, such as we have proposed.

*Testing for chaos*

Given our suggestion has superficial validity, patterns in existing data should be assessed to more robustly test the idea that chaotic dynamics underlie among-individual differences. One potential way to test for the presence of divergent trajectories from adjacent initial conditions is the calculation of Lyapunov exponents (LEs). LEs quantify the tendency for neighbouring trajectories to diverge or converge (Hastings et al., 1993). Divergent trajectories are then considered an indicator of sensitivity of initial conditions and therefore chaotic dynamics. LEs have been applied to various time series of animal populations to determine whether population sizes tend to fluctuate chaotically, and are therefore inherently unpredictable, or if they are unpinned by deterministic rules (e.g., Benincà et al., 2008). A review indicated that ecologists tend not to find positive LEs (Medvinsky et al., 2015), although periods of chaos amongst more stable dynamics are often observed (e.g., Becks & Arndt, 2008). These findings mirror our suggestion above that there may only be periods of ontogeny where the development of behaviour is chaotic, with other periods of time where behaviour develops more predictably. In such a scenario, where chaotic dynamics govern only part of behavioural ontogeny alongside more predictable process, signal and chaos may coexist and, as a result, it would be possible to detect repeatable behaviours, correlations with other traits such as life-history and fitness, as well as to detect genetic variance in behaviour among individuals; all of which occur in various studies throughout the personality literature. So, although our suggestion does not require these factors to be present, it is compatible with them being present alongside the chaotic development of behaviour.

Within the existing toolkit of behavioural ecologists, random regressions, also known as random slope models, can be used to determine whether among-individual differences are increasing or decreasing, or stable, over time (Brommer, 2013a, 2013b). In contrast to a typical linear model, where all members of the population are assumed to follow the same development trajectory, a random regression can fit a unique developmental trajectory for each individual. This allows the estimation of among-individual variance in both the height of these trajectories (their means or intercepts) and the slopes of gradients, as well as the covariance between an individual's slope and its intercept. In turn, one may then calculate the change in among-individual variance with age. If trajectories tend to diverge as we have suggested, the among-individual variance should increase with age. This pattern has indeed been observed in a study on wild field crickets (*Gryllus campestris*), where an increase in among-individual variance with age was detected for several behaviour traits (Fisher, David, Tregenza, & Rodriguez-Munoz, 2015). Bierbach et al. (2017) fitted observation number as a fixed effect, to determine that the mollies average exploration was increasing, as well as random intercepts for individuals, to demonstrate that individuals had different mean exploration levels. Neither of these, alone or in combination, are sufficient to determine whether or not among-individual differences are increasing through time. Individuals can increase, decrease, or show no change on average with an environmental factor and may independently diverge, converge or maintain parallel trajectories. While among-individual differences in intercepts indicate individuals have different mean behaviours, random slope models (or LEs) are needed to assess if individuals are becoming more or less different over the course of the experiment (Brommer, 2013a, 2013b; Fisher et al., 2015; Kluen & Brommer, 2013).

The experimental design required to test these ideas is not necessarily complicated, but the data requirements may still be quite challenging. Minimal environmental and genetic variation amongst the study organisms, as in Bierbach et al. (2017) and others, is desirable. Repeated behavioural assays, essentially starting at day zero, throughout the organisms' development would be required to track initial adjacent trajectories as they diverge (if this occurs). The recommended sample size for detecting among-individual variance in slopes is around 200 measures for random regression models, with the ratio of unique individuals to tests per individual around 2:1 (Martin, Nussey, Wilson, & Réale, 2011). To estimate LEs, sampling requirements are dependent on the number of parameters and length of the time series being analysed, but typically exceed >1,000 measurements (Eckmann & Ruelle, 1992; Wolf, Swift, Swinney, & Vastano, 1985).

While these requirements may seem formidable, such experiments could be made much more viable with automated tracking technology. If an organism was placed in a single container with food, water and shelter, and its natural movement tracked, for as long as it was in the container, then aspects of individual movement could be extracted at regular intervals, giving repeated measures of the same behaviour over extended periods of time. For instance, a simple experiment could be designed in which eggs or larvae of inbred fruit flies (*Drosophila* sp.) are collected shortly after laying and placed in individual vials containing sufficient resources for development and maturation. As the flies develop, individual behaviours could be continually measured using *Drosophila* activity monitors. These devices are commonly used to monitor either individual- or population-level locomotor activity and circadian rhythms in *Drosophila*, and present the possibility of simultaneously collecting data for 100-1,000s of individuals (Gilestro, 2012; Rosato & Kyriacou, 2006). Such continuous activity data could then be grouped into shorter intervals (e.g., 1–12 h), and analysed to assess the development and divergence of individual behaviour from the larval stage through to maturity. If similar assumptions regarding genetic and environmental variation are made to those in previous studies, such an experiment would allow one to test the hypothesis that small initial differences in phenotype among individuals lead over time, through chaotic dynamics, to consistent among-individual differences.

*Conclusions*

In summary, among-individual behavioural variation perhaps should be considered the null expectations in the majority of situations, rather than requiring any further variation in other traits such as life-history strategy. The recent rise in focus on consistent behavioural differences has demonstrated that it is present in essentially every laboratory and wild population. Therefore, the starting point to understanding behavioural variation should not be a model requiring equivalent variation from other sources, but perhaps one that incorporates non-linear and chaotic behavioural development. Embracing concepts from fields such as the study of complex systems, for example emergence and deterministic chaos, should give us a more realistic platform from which to understand variation in the natural world.

### Author contributions
All authors discussed the original manuscript and drafted the commentary.

### Competing interests
The authors declare no competing financial interests.

### Acknowledgements
We thank Kate Laskowski, Tom Sherratt and two anonymous referees for comments that greatly improved the clarity of our argument and the quality of the article in general. We also thank the University of Guelph's Graduate lounge, and Dougie and Pablo's, for providing reasonably priced pints that facilitated the initial discussion.